\documentclass[preprint,amsmath,amssymb,aps
]{revtex4-2}
\usepackage[english]{babel}
\usepackage[letterpaper,top=2cm,bottom=2cm,left=3cm,right=3cm,marginparwidth=1.75cm]{geometry}

\usepackage{amsmath,amssymb,cancel}
\usepackage{graphicx}
\usepackage[colorlinks=true, allcolors=blue]{hyperref}
\usepackage{bigstrut}
\def\bal#1\eal{\begin{align}#1\end{align}}
\def\be#1\ee{\begin{equation}#1\end{equation}}
\def\bsp#1\esp{\begin{split}#1\end{split}}
\newcommand{\brac}[1]{\left( #1 \right)}
\newcommand{\bfrac}[2]{\left( \frac{#1}{#2} \right)}
\newcommand{\eps}{\varepsilon}
\newcommand{\inv}[1]{\frac{1}{#1}}

\newcommand{\rd}{\mathrm{d}}
\newcommand{\re}{\mathrm{e}}
\newcommand{\ri}{\mathrm{i}}
\newcommand{\rF}{\mathrm{F}}
\newcommand{\rL}{\mathrm{L}}
\begin{document}
    
    \title{Constraining flavour-universal nonstandard interactions and superweak extension of the standard model}
    
    \author{Timo J.~K\"arkk\"ainen}
    \email{karkkainen@kbfi.ee}
    \affiliation{Laboratory of High Energy and Computational Physics,\\ National Institute of Chemical Physics and Biophysics, Rävala pst. 10, 10143 Tallinn, Estonia}
    \author{Zolt\'an Tr\'ocs\'anyi}
    \email{zoltan.trocsanyi@cern.ch}
    \affiliation{Institute for Theoretical Physics, ELTE E\"otv\"os Lor\'and University,
        P\'azm\'any P\'eter s\'et\'any 1/A, 1117 Budapest, Hungary\\
        ELKH-DE Particle Physics Research Group, 4010 Debrecen, PO Box 105, Hungary}
    
    \begin{abstract}
        Nonstandard neutrino interactions (NSI) arising from light 
        and heavy mediators probe different sectors 
        of the parameter space of models focusing on phenomena that 
        require the extension of the standard model. 
        High-energy scattering experiments are not relevant on 
        constraining the NSI hiding a light mediator at 
        the fundamental level, while flavour-universal NSI cannot be 
        probed with neutrino oscillation experiments. 
        Currently the only way to measure flavour-universal NSI with a 
        light mediator is to rely on coherent 
        elastic neutrino-nucleon scattering experiments, which we 
        use to derive bounds for light mediator 
        flavour-universal NSI. For light NSI, we obtain 
        $\eps^u \in [-14.85,14.79]$ and $\eps^d = [-13.19,13.84]$ (90~\% CL.). 
        We also derive constraints on flavour-universal heavy NSI and find a 
        2$\sigma$ tension. Finally, we discuss the implications of the 
        experiments on the 
        allowed parameter space of a specific example model, 
        called superweak extension of the standard model.
    \end{abstract}
    
    \keywords{gauged U(1) extension, nonstandard interactions, coherent elastic neutrino-nucleon scattering}
    \maketitle
    \thispagestyle{empty}
    \section{Introduction}
    
    The discovery of neutrino oscillations \cite{Fukuda:1998mi,Ahmad:2001an} kickstarted a plethora of 
    research efforts in neutrino physics. As the standard model 
    (SM) is devoid of neutrino masses, neutrinos are an exciting 
    option as a portal to new physics, which must contain a 
    mechanism to generate neutrino masses, and therefore neutrino 
    oscillations. One of the most popular models of mass generation 
    is the seesaw mechanism 
    \cite{Fritzsch75,Minkowski:1977sc,GellMann:1980vs,Yanagida:1979as,Mohapatra:1979ia,Mohapatra:1980yp,Schechter:1980gr,Magg:1980ut,Glashow:1979nm,Lazarides:1980rn,Gelmini:1980re,Foot1989}. The 
    type I mechanism introduces heavy  right-handed neutrinos that 
    are sterile under the SM. As at least two of the three active 
    neutrinos are massive, the minimum extension includes two 
    sterile neutrinos. In this paper we focus on the type I 
    mechanism; see Refs.~\cite{Mohapatra:2006gs,King:2014nza,King:2003jb} for 
    reviews on other types of neutrino mass generation mechanisms.
    
    New physics effects are manifested at low energy scales via 
    effective operators, which are generated by integrating out the 
    heavy degrees of freedom from the high-energy theory. In the 
    context of neutrino physics, there are three important 
    operators:
    \bal 
    \mathcal{O}_5 &= \frac{C_5}{\Lambda}(\overline{L^c} \cdot H)(H \cdot L)\,,\\
    \mathcal{O}_{6a} &= \frac{C_{6a}}{\Lambda^2}(\overline{L}\gamma^\mu P_\rL L)(\overline{f}\gamma_\mu P_X f)\label{eq:NSIoperator}\,,\\
    \mathcal{O}_{6b} &= \frac{C_{6b}}{\Lambda^2}(\overline{L}\cdot H)i\cancel{\partial}(H^\dagger\cdot L)
    \eal 
    where the dot represents the SU(2)$_\rL$ invariant product of 
    doublets and $\Lambda$ is the scale of new physics. The first 
    operator is the Weinberg operator \cite{Weinberg:1979sa}, which 
    is the only possible gauge invariant dimension-5 operator that 
    can be constructed from the SM fields. After spontaneous 
    symmetry breaking this gives a Majorana neutrino mass term. The 
    second operator corresponds to \textit{nonstandard interactions}
    (NSIs) \cite{Grossman:1995wx} of four charged leptons or 
    charged lepton -- quark NSI that in general break flavour. The 
    third operator arises from active-sterile neutrino mixing. The 
    latter two operators are of dimension six.
    
    The scale $\Lambda$ is interpreted as the energy scale of new 
    physics, typically considered much higher than the electroweak 
    scale, corresponding to a {\em heavy NSI mediator} at the 
    fundamental level. This expectation is based on the assumption 
    that the couplings $C_i$ are $\mathcal{O}(1)$ coefficients. 
    However, quantum field theory does not \textit{a priori} force 
    the couplings to be so large. In the SM, a prime example of 
    small couplings is the Yukawa coupling of the electron, 
    $y_e \simeq 3 \cdot 10^{-6} \ll 1$. In the case when the 
    couplings $C_i \ll 1$, the scale $\Lambda$ can be as low as GeV 
    or even MeV, and the mass of the corresponding NSI mediator may 
    be light or similar when compared to the momentum transfer in 
    the experiment. While such scenarios do not support models 
    built on naturalness arguments, they are certainly not ruled 
    out, and also predictive. Such new physics interactions can be 
    probed at high-intensity, low-energy experiments that are 
    planned for the next decades, at the European Spallation 
    Source \cite{Baxter:2019mcx} (ESS) for example. The ESS 
    promises an order of magnitude increase in neutrino flux as 
    compared to that of the Spallation Neutron Source where the 
    first successful detection of coherent elastic neutrino-nucleon 
    scattering (CE$\nu$NS) \cite{Freedman:1973yd} was carried out by the COHERENT 
    experiment  \cite{COHERENT:2017ipa,COHERENT:2020iec}. The 
    increase in statistics is the key to improve the bounds on the NSI parameters.

    Neutrino interactions have very low cross sections. Nonetheless 
    neutrino-electron and neutrino-nucleon cross sections have been 
    measured at scattering experiments where the averaged momentum 
    transfer squared is large, $\langle q^2 \rangle = 20$\,GeV$^2$ 
    \cite{CHARM:1986vuz,NuTeV:2001whx,LSND:2001akn}. These 
    measurements give stringent bounds to new physics effects 
    originating from the effective operators, namely the NSI with 
    new physics scale $\Lambda > \Lambda_\text{EW}$. 
    The observation of CE$\nu$NS by COHERENT allows us to 
    test whether or not NSI effects exist with scale $\Lambda$ 
    significantly below the electroweak scale.
    
    Different extensions of the SM produce different NSI textures. 
    A subclass of these extensions is flavour conserving. 
    Consequently, the NSI matrix is diagonal and real, containing 
    only three elements, which have contributions from up-type 
    quarks, down-type quarks and charged leptons. If in addition 
    the extension is flavour universal, then the NSI matrix is 
    isotropic (proportional to the unit matrix). In the bottom-to-
    top approach, current experimental bounds can be used to 
    constrain the high-energy theory parameters. In contrast, the 
    top-to-bottom approach can be used to predict the texture and 
    region NSI available for a particular UV complete model.
    
    In this paper, we discuss the NSI formalism and both approaches 
    by considering the constraints with light and heavy NSI 
    mediators. We derive bounds for flavour-universally coupled NSI 
    mediator in {\em both the light and the heavy case}. We 
    consider a specific example, the super-weak extension of the 
    standard model (SWSM) \cite{Trocsanyi:2018bkm} which exhibits tiny flavour-universal 
    couplings to fermions. A similar study but in the context of 
    different models has been recently carried out \cite{Denton:2018xmq}. 
    The SWSM contains an NSI mediator that is light in scattering 
    experiments and therefore it evades detection, but not so in 
    CE$\nu$NS, which is sensitive for NSI originating from SWSM. We 
    derive bounds on the new gauge coupling and ratio of the vacuum 
    expectation values in the SWSM based on the results of COHERENT 
    \cite{COHERENT:2017ipa,COHERENT:2020iec,Giunti:2019xpr} and our 
    previous analyses on dark matter \cite{Peli:2022ybi} in the SWSM.
    
    Our paper is organized as follows. We introduce the NSI formalism in Section II and present the bounds for flavour-universal NSI parameters in the case of light and heavy mediator. We then discuss in Section III a particular example of a model which exhibits flavour-universal NSI --- the superweak extension of the standard model. In Section IV we present our results: on one hand, the bounds from COHERENT constrain the gauge parameters of the superweak extension, while on the other the properties of the model predict texture and correlations for the NSI parameters. We present our conclusions in Section V.

    \section{Experimental constraints on the NSI parameters}
    
    \subsection{NSI formalism}
    
    In our study we focus on the ${\cal O}_{6a}$ operator of 
    Eqn.~\eqref{eq:NSIoperator} that is relevant to neutrino-matter 
    interactions. In the usual parametrization of the NSI 
    Lagrangian the interaction strength is set by the Fermi 
    coupling $G_\rF$,
    \be
    {\cal L}_{\rm NSI} = -2\sqrt{2} G_\rF \sum_{f,X=\pm,\ell,\ell'} \eps^{f,X}_{\ell,\ell'} (\bar{\nu}_\ell \gamma^\mu P_\rL \nu_{\ell'})(\bar{f} \gamma_\mu P_X f)
    \label{eq:LNSI}
    \ee
    where $\eps^{f,X}_{\ell,\ell'}$ parametrizes the strength of 
    the new interaction with respect to $G_\rF$, with $\ell$, 
    $\ell'$ denoting charged lepton flavours and $f$ being a 
    charged fermion in the standard model.
    
    When one matches the NSI Lagrangian \eqref{eq:LNSI} with the 
    effective Lagrangian obtained from a high-energy theory, the NSI
    parameters are proportional to the propagator of the mediator, 
    i.e.~to $\eps^{f,X}_{\ell,\ell'} \propto (q^2-M^2)^{-1}$, where 
    $q^\mu$ is the four-momentum ($q^2 = q_\mu q^\mu$) carried by 
    the mediator and $M$ is its mass. In a neutrino scattering 
    experiment, we may approximate the propagator either as
    \be 
    \eps^{f,X}_{\ell,\ell'}
    \propto
    +\inv{q^2}\text{ if }q^2 \gg M^2,
    \label{eq:lightNSI}
    \ee 
    or
    \be 
    ~\eps^{f,X}_{\ell,\ell'} \propto 
    -\inv{M^2}\text{ if }q^2 \ll M^2. 
    \label{eq:heavyNSI}
    \ee 
    The first case in Eq.~\eqref{eq:lightNSI} corresponds to 
    ``light NSI mediator'', while the second one to ``heavy NSI 
    mediator''. For concreteness, let us consider $M=50$ MeV. Then 
    the mediator is considered {\em heavy} from the {\em viewpoint 
        of neutrino oscillation experiments}, but {\em light for high-
        energy neutrino scattering experiments}, such as CHARM 
    \cite{CHARM:1986vuz} and NuTeV \cite{NuTeV:2001whx}. However, 
    if $q^2$ is similar in size to $M^2$, as in the case of 
    CE$\nu$NS in our example, we cannot take any of these limits. 
    Nevertheless, we can still apply the NSI formalism using the 
    full propagator with $q^2$ being the characteristic momentum 
    transfer squared in the scattering experiment. The resulting NSI
    couplings interpolate smoothly between the light and heavy 
    limits. We present an example in Sect.~\ref{sec:SWSM}.
    
    \subsection{Global fit of the heavy NSI parameters}
    
    Most studies reporting the bounds on NSI parameters simplify 
    the fitting procedure to contain only one nonvanishing NSI 
    parameter. For flavour-universal models these bounds can not be 
    considered, and a global fit instead is required. This has been 
    performed in Refs.~\cite{Coloma:2017egw,Giunti:2019xpr} for example.
    
    The $\chi^2$-fitting in \cite{Giunti:2019xpr} follows the standard procedure, where the function
    \be 
    \chi^2 = \sum \limits_i \bfrac{N_i^\text{exp}-(1+\alpha_c)N_i^\text{theor}+(1+\beta_c)B_i}{\sigma_i}+\bfrac{\alpha_c}{\sigma_{\alpha_c}}^2+\bfrac{\beta_c}{\sigma_{\beta_c}}^2+\bfrac{\eta_c-1}{\sigma_{\eta_c}}^2
    \ee 
    is minimized with respect to the nuisance parameters $\alpha_c$, $\beta_c$ 
    and $\eta_c$, corresponding to the systematic uncertainties due to the 
    signal rate, the background rate and a quenching factor. The factors 
    $\sigma_{\alpha_c}$, $\sigma_{\beta_c}$, $\sigma_{\eta_c}$ are the 
    corresponding standard deviations, and the $\sigma_i$ are the 
    uncertainties of the number of events in an energy bin 
    $N_i^\text{exp}$ from Ref.~\cite{COHERENT:2017ipa},
    and $N_i^\text{theor}$ is the theory estimate in the same bin.
    
\begin{figure}[t]
        \centering
        \includegraphics[width=\linewidth]{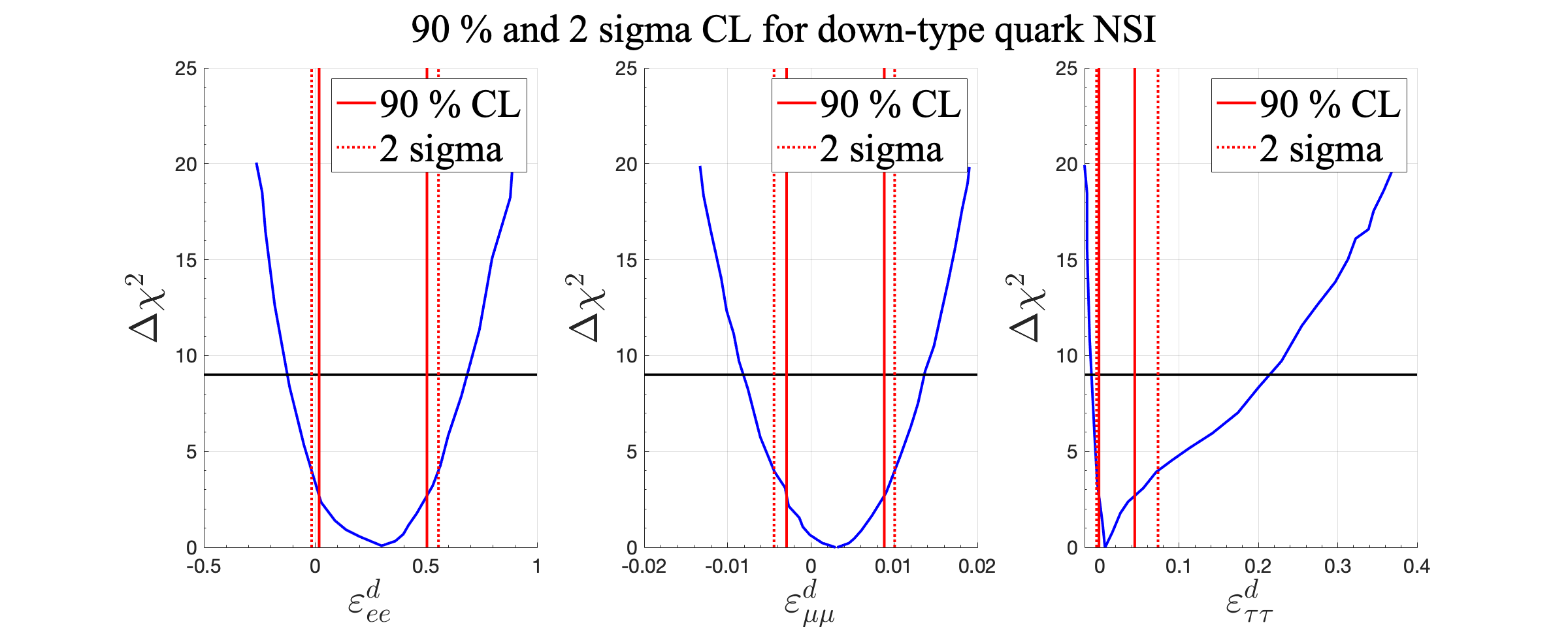}
        \includegraphics[width=\linewidth]{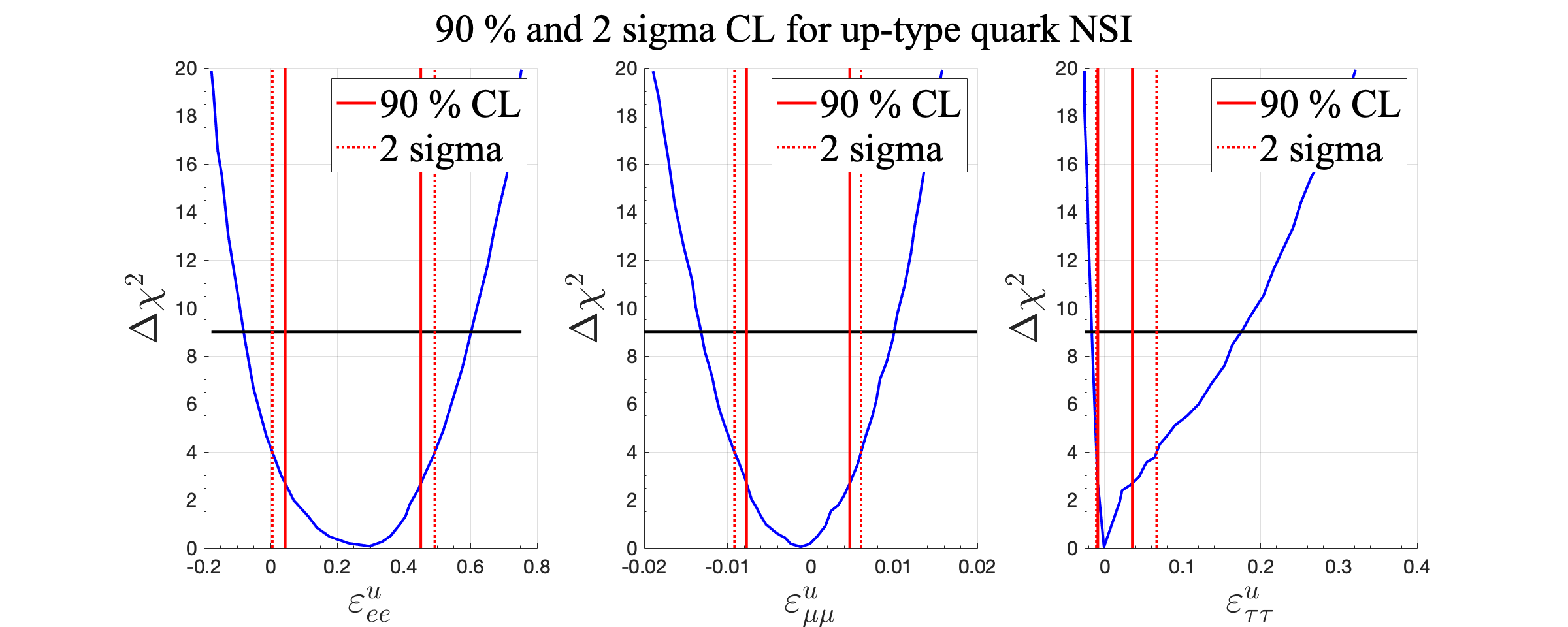}
        \caption{Determinations of 2$\sigma$ and 90 \% confidence intervals from minimized $\Delta \chi^2$-distributions given in \cite{Coloma:2017egw}. Down-type quark NSI above and up-type quark NSI below. 
            The vertical black line $(\Delta \chi^2 = 9)$ corresponds to the $3\sigma$ bound, used to find the values in Table \ref{tab:NSIbounds3}.
            \label{fig:chi2}}
    \end{figure}

    In Ref.~\cite{Coloma:2017egw} the authors perform a global fit 
    to current experiments for the NSI couplings with heavy 
    mediators and in the absence of CP violation, that is, the NSI 
    parameters are assumed to be real. The authors performed a 
    $\chi^2$-test (their procedure for COHERENT assumes only 
    one nuisance factor, and the analysis predates the start of 
    COHERENT experiment), minimizing the $\chi^2$-function, and 
    presented the dependence of the $\Delta \chi^2$-distributions 
    (the difference of a $\chi^2$-test value to $\chi^2$ best-fit 
    value), that is, the statistical significance of the NSI 
    parameters. We reproduced those plots here in 
    Fig.~\ref{fig:chi2}, with $2\sigma$ and 90\,\% confidence 
    intervals exhibited. We read off the best-fit points directly 
    from these graphs, and presented those together with the 
    confidence intervals in Table~\ref{tab:NSIbounds2}.
    \begin{table} \centering 
        \begin{tabular}{|c||c|c||c|c|}\hline 
            \textbf{Parameter}  & \textbf{Best-fit point} $\mu_i$ & \textbf{$2\sigma$ CI} $\sigma_{2,i}$ & \textbf{90 \% CI} $\sigma_{90,i}$ \\\hline \hline 
            $\eps^d_{ee}$       & 0.301    & [--0.015, 0.556] & [0.019, 0.504]  \bigstrut \\\hline 
            $\eps^d_{\mu\mu}$   & 0.003    & [--0.004, 0.010] & [--0.003, 0.009] \bigstrut \\\hline 
            $\eps^d_{\tau\tau}$ & 0.006    & [--0.004, 0.073] & [--0.001, 0.044] \bigstrut \\\hline 
            $\eps^u_{ee}$       & 0.297    & [0.006, 0.493] & [0.044, 0.451]  \bigstrut \\\hline 
            $\eps^u_{\mu\mu}$   & $-$0.001 & [--0.009, 0.006] & [--0.008, 0.005] \bigstrut \\\hline 
            $\eps^u_{\tau\tau}$ & $-$0.001 & [--0.011, 0.067] & [--0.009, 0.035] \bigstrut \\\hline 
        \end{tabular}
        \caption{\label{tab:NSIbounds2}Best-fit points for diagonal quark NSI parameters, and also 90 \% and $2\sigma$ confidence intervals (CI) derived from using Fig.~4 of \cite{Coloma:2017egw}. The bounds apply only for heavy mediator NSI ($M^2 \gg 20$\,GeV$^2$).}
    \end{table} 
    
\begin{figure}[t]
        \centering
        \includegraphics[width=0.49\linewidth]{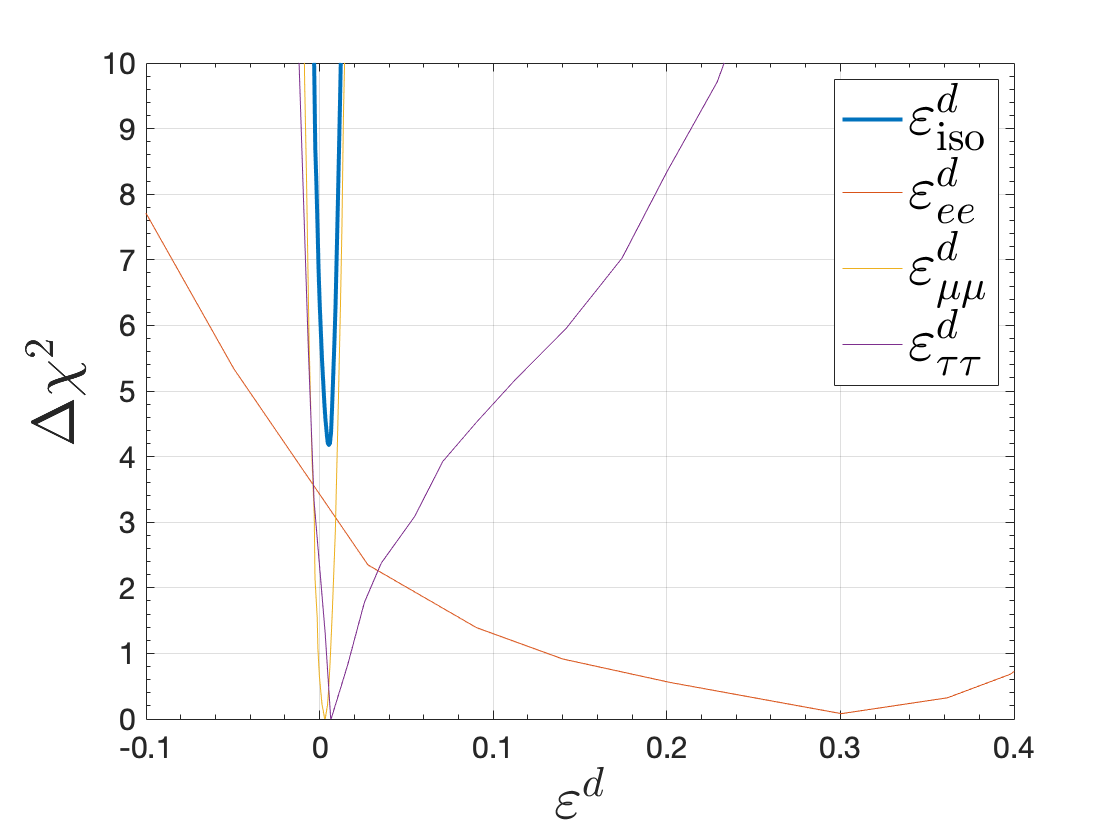}
        \includegraphics[width=0.49\linewidth]{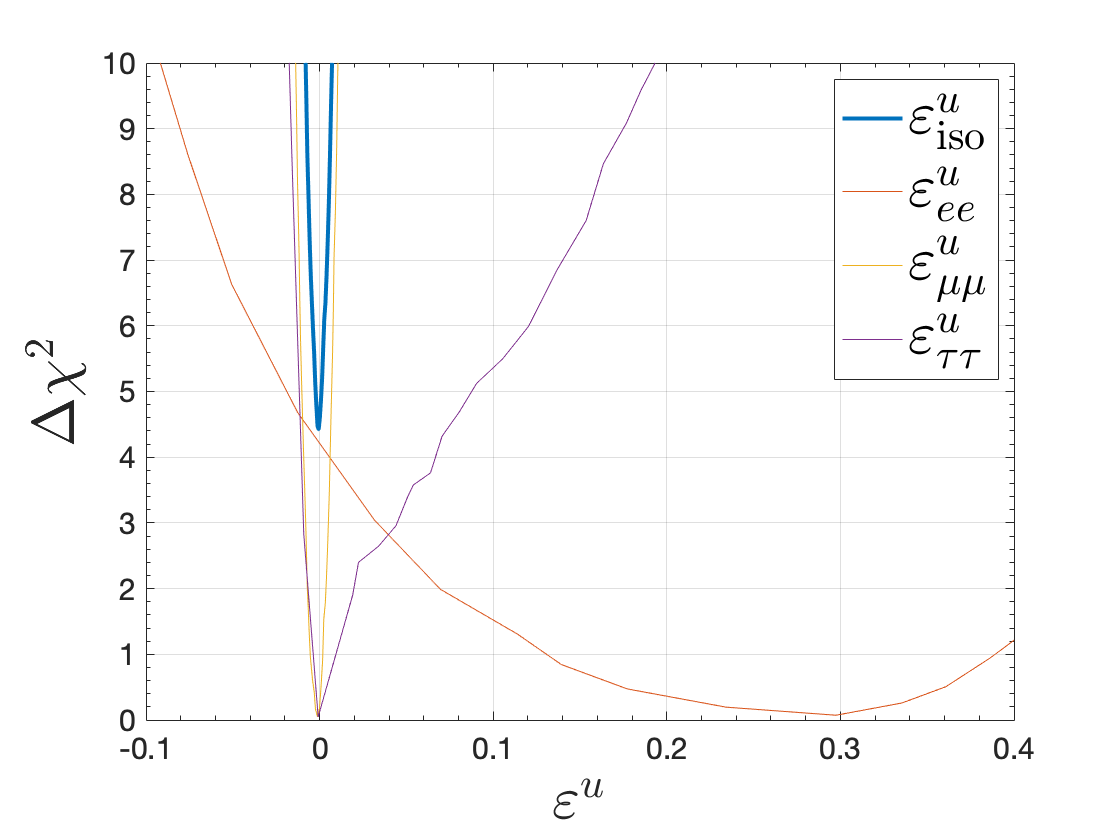}
        \caption{ Combined $\chi^2$-distributions and the individual components overlaid. 
            \label{fig:combined-chi2}}
    \end{figure}

\begin{figure}[ht]
        \centering
        \includegraphics[width=0.49\linewidth]{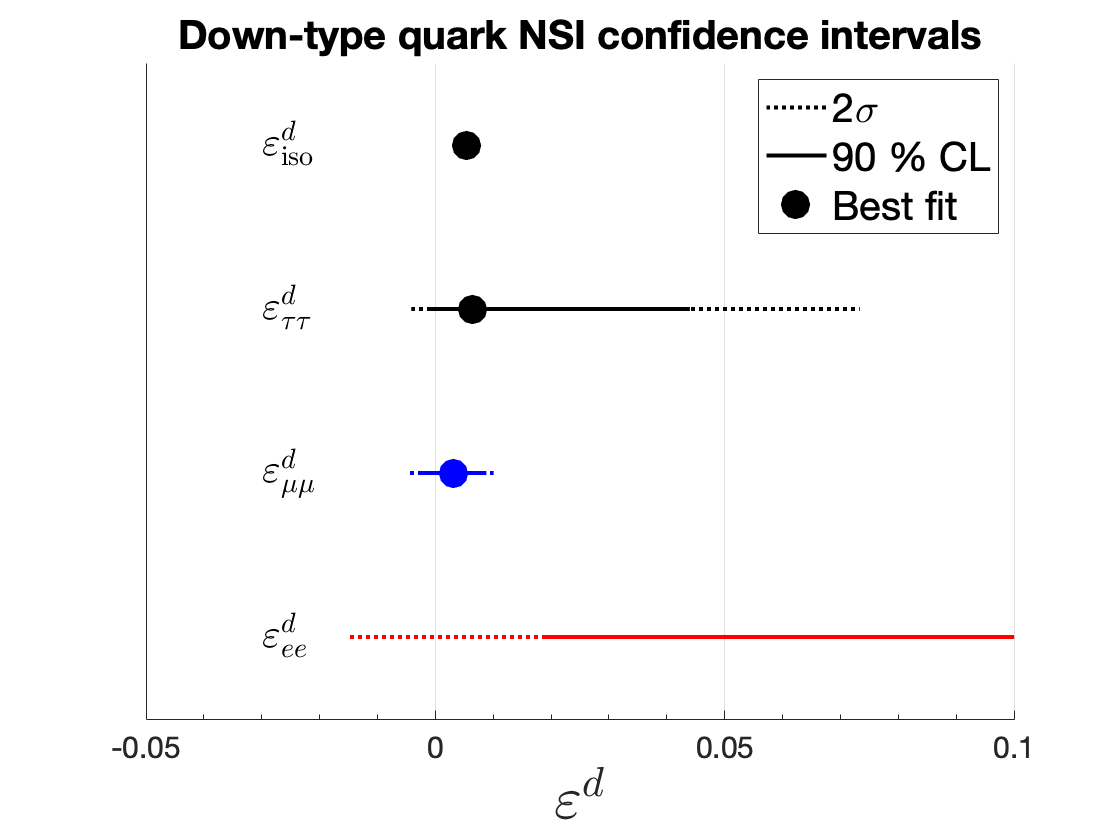}
        \includegraphics[width=0.49\linewidth]{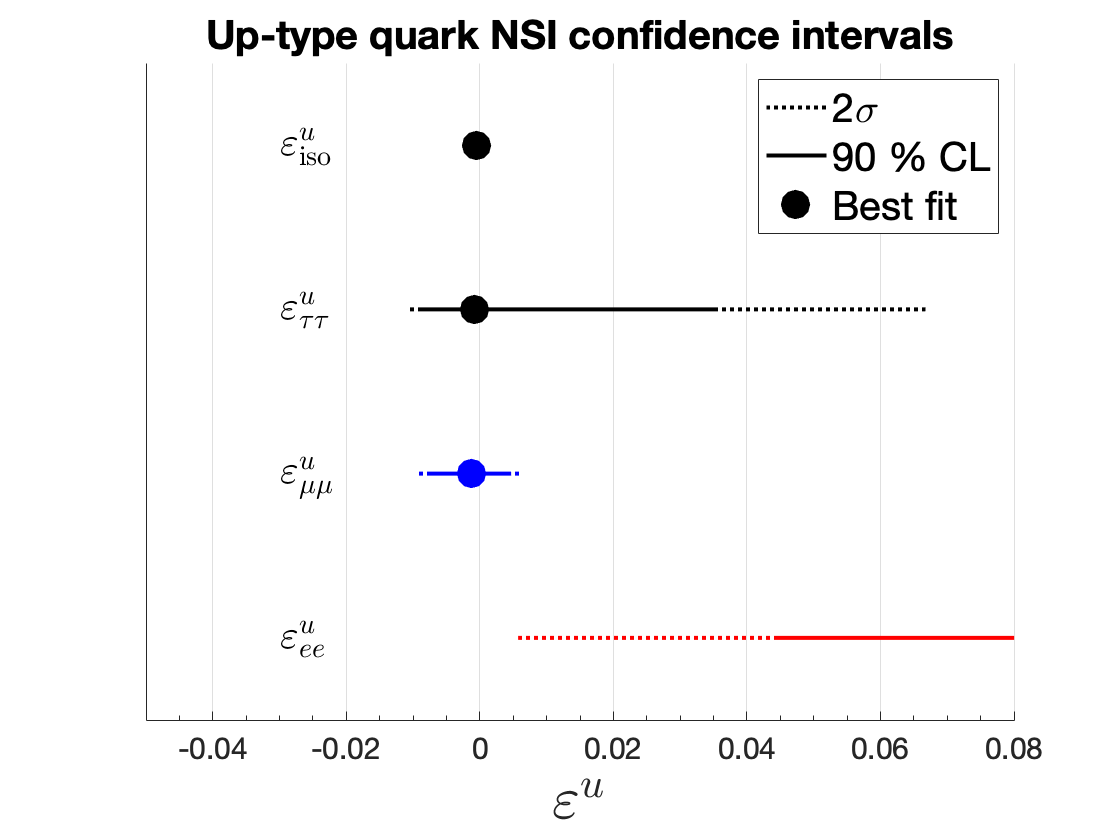}
        \caption{ Comparisons of 2$\sigma$ and 90 \% confidence intervals for the diagonal elements, including best-fit value. Left: down-type quark NSI, right: up-type quark NSI. Isotropic NSI included. The best-fit of $\eps_{ee}$ is not visible at this range.
            \label{fig:NSI-intervals}}
    \end{figure}

    We then combined the individual $\Delta\chi^2$-distributions 
    to test flavour-universal couplings by summing the three 
    $\Delta\chi^2$-distributions \cite{Koziol1978}:
    \be 
    \Delta\chi^2_\text{isotropic} =
    \Delta\chi^2_{ee} + \Delta\chi^2_{\mu\mu} + \Delta\chi^2_{\tau\tau}
    \,.
    \ee
    We present the combined up- and down-type isotropic heavy NSI 
    coupling $\Delta\chi^2$-distributions in 
    Fig.~\ref{fig:combined-chi2}, with the individual original 
    distributions overlaid. The relative incompatibility of 
    different flavour distributions results in tension with 
    experimental data indicating that both the up- and down-type 
    quark isotropic NSI scenario are excluded at $2\sigma$. 
    We compare the individual and combined bounds in Fig.~\ref{fig:NSI-intervals}.
    For isotropic NSI we have summarized our results in Table \ref{tab:NSIbounds3}.
    These bounds are relevant for theories which are accessible 
    via high-energy experiments, where the mediator has at least a 
    mass of $\mathcal{O}(10)$\,GeV and couples to quark flavours universally.
    \begin{table}[ht] \centering 
        \begin{tabular}{|c||c|c|c|}\hline 
            \textbf{Parameter} & \textbf{Best fit} & $3\sigma$ \textbf{CI} \\\hline 
            $\eps^u$ & $-5.5\times 10^{-4}$ & $[-0.0073,0.0063]$ \bigstrut \\\hline 
            $\eps^d$ & $5.3 \times 10^{-3}$ & $[-0.0026,0.0114]$\bigstrut \\\hline 
        \end{tabular}
        \caption{\label{tab:NSIbounds3} Best-fit points and $3\sigma$ confidence intervals for isotropic NSI. The constraints from high-energy experiments have been taken into account, hence the bounds apply only for heavy mediator NSI ($M^2 \gg 20$ GeV$^2$).}
    \end{table}
    
    For leptonic NSI, one can use the constraints given in Fig.~2 
    of Ref.~\cite{Barranco:2007ej}, where the authors performed 
    both one-parameter- and flavour-conserving fits. Their 
    $\chi^2$-analysis takes into account the data from LEP 
    experiments (ALEPH, DELPHI, L3 and OPAL), LSND experiment, 
    reactor experiments (MUNU and Rovno) and CHARM II experiment. 
    The Borexino experiment has performed one-parameter fits 
    \cite{Borexino:2019mhy} leading to the loosest bound. 
    
    \subsection{Flavour universal NSI from the COHERENT experiment}
    
    For obtaining constraint on light NSI parameters oscillation 
    experiments can be utilized. However, those cannot observe the 
    diagonal elements of the NSI matrix themselves. Instead, they 
    measure off-diagonal couplings and differences of the diagonal 
    couplings. In Ref.~\cite{Biggio:2009nt} the authors have 
    chosen the convention that $\eps_{\mu\mu}$ is subtracted from 
    the effective Mikheyev-Smirnov-Wolfenstein neutrino 
    oscillation Hamiltonian as a phase rotation, so the observable 
    parameters are $\eps^f_{ee} - \eps^f_{\mu\mu}$ and 
    $\eps^f_{ee} - \eps^f_{\tau\tau}$. Consequently,  
    flavour-conserving NSI (that is, diagonal NSI matrix) can be 
    detected in neutrino oscillations only if it is not flavour-
    universal. In flavour-universal case the NSI matrix is 
    isotropic and manifests itself as an unphysical phase 
    rotation, undetectable in such experiments. 
    
    Another resource to test the light NSI couplings is coherent 
    elastic neutrino-nucleon scattering (CE$\nu$NS). In this 
    experiment the differential cross section in the recoil energy 
    $T$ ($T \lesssim 10$\,keV) of the nucleus in this process is given by
    \be 
    \frac{\rd\sigma}{\rd T} = \frac{G_\rF^2M}{\pi}\brac{1-\frac{|\textbf{q}|^2}{4E_\nu^2}}Q_W^2
    \ee 
    where $M$ is the mass of the nucleus and $|\textbf{q}|^2 =2MT$ 
    is the momentum transfer squared. $E_\nu$ is the energy of the 
    neutrino, while $Q_W$ denotes the weak charge for a nucleus of 
    $Z$ protons and $N$ neutrons, which in the standard model reads as
    \be 
    Q_W^\text{SM} = g_V^n N F_n(\textbf{q})+g_V^p Z F_p(\textbf{q})
    \,,\quad 
    g_V^n = -\frac{1}{2},\quad g_V^p = \frac{1}{2}-2\sin^2\theta_W
    \,.
    \label{eq:QWSM}
    \ee 
    The functions $F_n$ and $F_p$ are nuclear form factors for the neutron and the proton distribution
    in the nucleus, parameterized using Helm's parameterization in Ref.~\cite{Giunti:2019xpr}:
    \be 
    F_x(|\textbf{q}|) = \frac{3j_1(|\textbf{q}|R_{x,0})}{|\textbf{q}|R_{x,0}}
    \re^{-|\textbf{q}|^2s^2/2}
    ,\quad 
    R_{x,0}^2 = 5 s^2 - \frac{5}{3} R_x^2
    ,\quad
    x=n\text{~or~} p
    \,.
    \ee 
    In this formula $R_{x,0}$ is obtained using the surface thickness $s = 0.9$\,fm and the 
    the root mean square radii of the proton and neutron distributions inside the nucleus.
    For instance, 
    $R_p(^{133}Cs) = 4.804$\,fm and $R_n(^{133}Cs) = 5.01$ for Cesium 
    and 
    $R_p(^{127}I) = 4.749$\,fm and $R_n(^{127}Cs) = 4.94$ for Iodine used in the experiments.
    The function $j_1(x) = \frac{\sin x}{x^2}-\frac{\cos x}{x}$
    is the spherical Bessel function of the first kind, order 1.
    
    CE$\nu$NS was predicted by Freedman in 1974 \cite{Freedman:1973yd}, 
    and finally observed for the first time in COHERENT experiment 
    in 2017 \cite{COHERENT:2017ipa}. The first run used Cesium-133 
    and Iodine-127 nuclei in 2017 and the second run liquid Argon-
    40 in 2020\cite{COHERENT:2020iec}.
    
    The generalization of the weak charge in Eq.~\eqref{eq:QWSM} to the case of generic NSI is
    \be\bsp
    Q_{W,e}^2 &=
    \brac{(g_V^p + 2\eps_{ee}^{u} + \eps_{ee}^{d}) Z F_p(|\textbf{q}|)
        +(g_V^n+\eps_{ee}^u + 2\eps_{ee}^d) N F_n(|\textbf{q}|)}^2\\
    &+ \left| (2\eps_{e\mu}^u + \eps^d_{e\mu}) Z F_p(|\textbf{q}|) 
    + (\eps_{e\mu}^u+2\eps_{e\mu}^d) N F_n(|\textbf{q}|))\right|^2\\
    &+ \left| (2\eps_{e\tau}^u + \eps^d_{e\tau}) Z F_p(|\textbf{q}|) + (\eps_{e\tau}^u+2\eps_{e\tau}^d) N F_ (|\textbf{q}|))\right|^2
    \esp
    \label{eq:QW2}
    \ee
    where $\eps^{f}_{\ell\ell'} = \eps^{f,+}_{\ell\ell'} + \eps^{f,-}_{\ell\ell'}$.
    
    
    We remark that the leading order contribution to the flavour-
    breaking NSI parameters $\eps_{\ell\ell'}^f$ ($\ell \neq \ell'$) 
    is proportional to the second order of those parameters, while 
    the flavour-conserving parameters contribute at both first and 
    second order (linear and square terms). If both flavour-
    conserving and flavour-breaking NSI parameters have 
    approximately the same magnitude and are significantly less 
    than one, then we may neglect the second order terms. Then, 
    the flavour-conserving NSI parameters dominate the distortion 
    to the weak charge $Q_W^2$:
    \be\bsp
    &Q_{W,e}^2 = Q^\text{SM}_{W,e} 
    \\ &\quad
    + 2 (g_V^n)^2 \Big(\eps^u_{ee} + 2\eps^d_{ee}\Big)N^2 F_n^2 
    + 6 g_V^n g_V^p \Big(\eps^u_{ee} +\eps^d_{ee}\Big) N Z F_n F_p 
    + 2 (g_V^p)^2 \Big(2\eps^u_{ee} + \eps^d_{ee}\Big) Z^2 F_p^2
    \,.
    \esp\ee
    Presently large values (larger than one) for the light NSI 
    parameters are still allowed experimentally for both flavour-
    conserving and flavour-breaking case \cite{Giunti:2019xpr}. In 
    such a case, one should use the complete formula for the weak 
    charge as given in Eq.~\eqref{eq:QW2}.
    
    We may utilize the COHERENT  limit given by \cite{Giunti:2019xpr} to constrain $\eps^q_{ee}$. 
    Analogously, the same argument can be used to demonstrate the dominance of the $\mu\mu$ elements on $Q_{W,\mu}^2$.
    
\begin{figure}[t]
        \centering
        \includegraphics[width=0.49\linewidth]{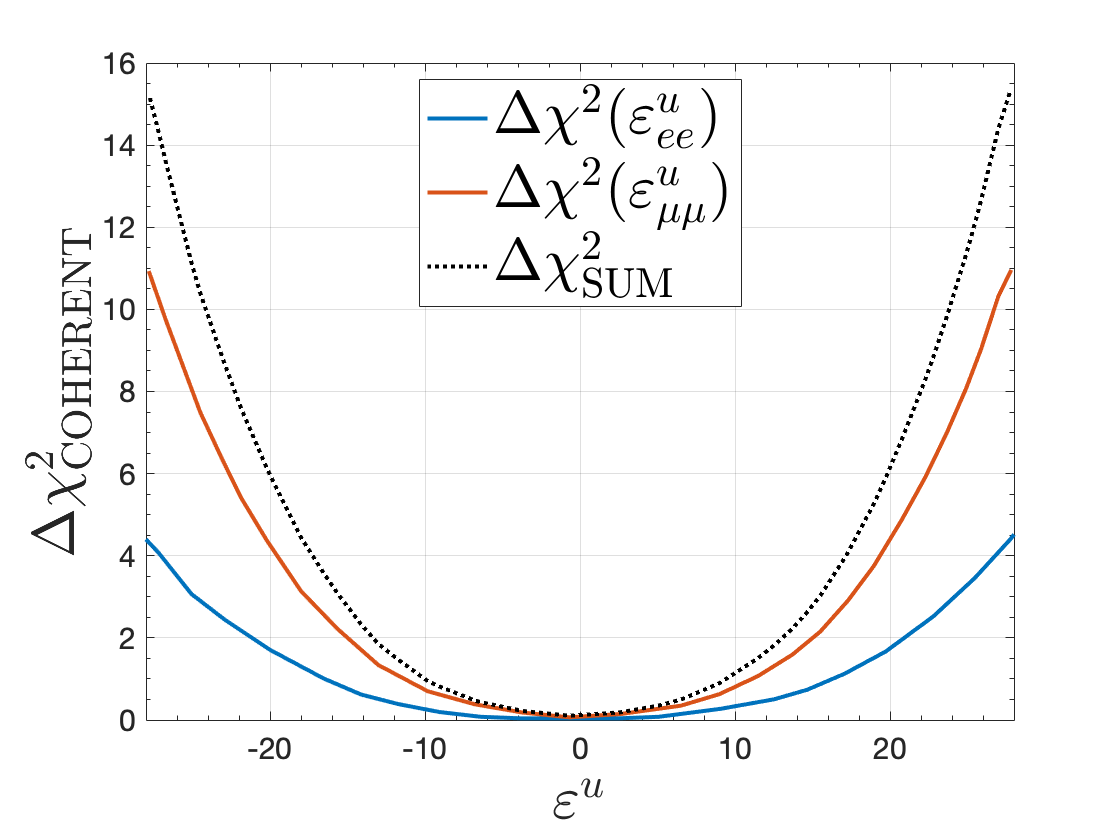}
        \includegraphics[width=0.49\linewidth]{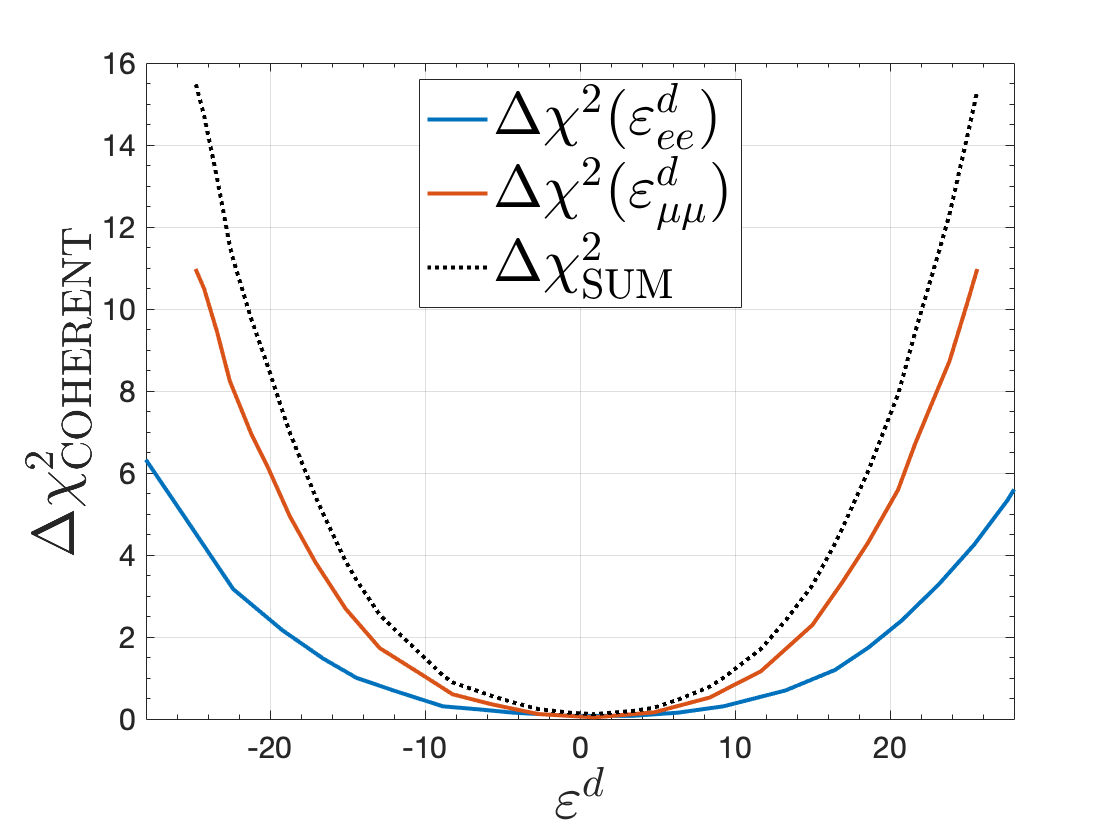}
        \caption{Combined $\Delta\chi^2$-distributions and the individual components overlaid. Only COHERENT data is taken into account.
            \label{fig:chisum}}
    \end{figure}

    
    We performed the combination of $\Delta\chi^2$-distributions 
    also for COHERENT experiment, which is sensitive to 
    $\eps_{ee}^q, \eps_{e\mu}^q$ and $\eps_{\mu\mu}^q$ but not to 
    $\eps_{\tau\tau}^q$, where $q=u,d$. In isotropic NSI models 
    $\eps_{ee}^q = \eps_{\mu\mu}^q$. We assume the COHERENT 
    measurements of these two couplings to be independent and sum 
    the $\Delta\chi^2$-distributions related to these parameters, 
    following the instruction of Ref.~\cite{Koziol1978}. We then 
    derive the COHERENT bounds for isotropic NSI parameters. We 
    reproduce the individual $\Delta\chi^2$-distributions taken 
    from Ref.~\cite{Giunti:2019xpr}, and show them together with 
    the combination in Fig.~\ref{fig:chisum}. The corresponding 
    confidence intervals are given in Table~\ref{tab:NSIbounds4}.
    
    \begin{table}[h] \centering 
        \begin{tabular}{|c||c|c|c|}\hline 
            \textbf{Parameter} & $2\sigma$ \textbf{CI} & \textbf{90 \% CI} & $1\sigma$ \textbf{CI} \\\hline\hline 
            $\eps^u$ & $[-17.25, 17.16]$ & $[-14.85,14.79]$ & $[-10.01,9.42]$\bigstrut \\\hline 
            $\eps^d$ & $[-15.31, 16.05]$ & $[-13.19, 13.84]$ & $[-8.61,9.23]$\bigstrut \\\hline 
        \end{tabular}
        \caption{\label{tab:NSIbounds4} Confidence intervals for isotropic NSI couplings based on the COHERENT constraints.}
    \end{table}
    
    A somewhat similar analysis in Ref.~\cite{Coloma:2017ncl} has 
    provided bounds for both diagonal and non-diagonal NSI 
    couplings for up- and down-type quarks. It has been completed 
    with the assumption that NSI coupling is the same for both 
    oscillation and CE$\nu$NS experiments. This is true if the 
    momentum transfer $q^2 \ll M^2$, where $M$ is the NSI
    mediator mass (heavy NSI mediator). However, in the case of 
    light NSI this condition does not hold. Our analysis is performed in the 
    region where the NSI coupling is $g^2/(q^2 - M^2)$, in contrast 
    to oscillations, where the coupling is $g^2/q^2$.
    
    \section{NSI couplings derived in the SWSM}
    
    \begin{table}[]
        \centering
        \renewcommand{\arraystretch}{1.3}
        \begin{tabular}{|c||c|c|c|c|c|c|}\hline 
            \textbf{Field}  & $Q_L$ & $u_R$ & $d_R$ & $L_L$ & $\ell_R$ & $N_R$\\\hline \hline 
            U(1)$_z$ charge & $\dfrac{1}{6}$ & $\dfrac{7}{6}$ & $-\dfrac{5}{6}$ & $-\dfrac{1}{2}$ & $-\dfrac{3}{2}$ & $\dfrac{1}{2}$ \bigstrut \\\hline 
        \end{tabular}
        \caption{Charges of the extra U(1) symmetry of the fermions in SWSM.
            \label{tab:charges}}
    \end{table}
    
    In this section we provide an example of a model that  naturally yields an isotropic NSI matrix, namely, the super-weak extension of the standard model \cite{Trocsanyi:2018bkm}. We recall the details of the SWSM only to the extent needed to derive the NSI couplings. For more details on the model, we call attention to Refs.~\cite{Iwamoto:2021fup,Peli:2022ybi,Iwamoto:2021wko,Karkkainen:2021tbh} where various phenomenological aspects were studied. 
    
    \subsection{Super-weak extension of the standard model}
    
    The SWSM is based on the SU(3)$_c\otimes$SU(2)$_L\otimes$U(1)$_Y\otimes$U(1)$_z$ gauge group. The U(1) gauge couplings are denoted by $g_y$ and $g_z$. 
    The anomaly-free U(1)$_z$ charges for the fermions are presented in Table~\ref{tab:charges}. The SU(2)$_L\otimes$U(1)$_Y$ symmetry is broken by the vacuum expectation value $v$ of the usual Brout-Englert-Higgs field, while the U(1)$_z$ symmetry is spontaneously broken by the vacuum expectation value $w$ of a complex scalar singlet (under transformations of the SM), making the corresponding neutral gauge bosons $Z$ and $Z'$ massive. These bosons mix weakly with mixing angle $\theta_Z$. 
    
    The covariant derivative related to the Abelian sector of the model is
    \be 
    D_\mu \supset D_\mu^{\rm U(1)} = \partial_\mu - \ri (y,z)\begin{pmatrix}
        g_y & -\eta g_z \\ 0 & g_z
    \end{pmatrix} R_\eps\binom{B_\mu}{B_\mu'} 
    \ee 
    where $R_\eps$ is an unphysical rotation matrix (whose rotation angle can be absorbed in $\theta_Z$), $y$ and $z$ are the U(1) charges, and the parameter $\eta$ is a convenient way to parametrize the kinetic mixing between the U(1) gauge fields. It depends on the renormalization scale scale $\mu$ mildly, and its value at the electroweak scale will vary according to the free choice of the scale $\mu_0$ where the mixing vanishes, $\eta(\mu_0)=0$. For $\mu_0$ chosen in the range $[M_Z, M_\mathrm{GUT}]$ one finds $\eta(M_Z) \in [0,0.656]$ \cite{Iwamoto:2021fup}. The largest value corresponds to a special case, where we assume that the kinetic mixing vanishes near the Planck scale.
    
    The interaction vertices can be obtained using the implementation of the model \cite{Iwamoto:2021wko} in \texttt{SARAH} \cite{Staub:2009bi,Staub:2010jh,Staub:2013tta}. For the $Z'$-neutrino interactions, we find
    \bal 
    -\ri eC^L_{Z'\nu_i\nu_k} 
    = -\frac{\ri }{2} &\Big[\sum \limits_{j=1}^3(\textbf{U}_{i,j})(\textbf{U}^\dagger)_{j,k} \left(\frac{e}{\sin \theta_W\cos \theta_W} \sin \theta_Z + (\eta -1) g_z \cos \theta_Z \right)\\
    &-g_z \cos \theta_Z \sum\limits_{j=1}^3\textbf{U}_{i,j+3} (\textbf{U}^\dagger)_{j+3,k}\Big]
    \eal 
    where $\theta_W$ is Weinberg's angle and \textbf{U} is the neutrino mixing matrix. The model contains three extra heavy sterile right-handed neutrinos $N_{R,i}$ ($i = 1$, 2, 3), so this matrix is a $6\times 6$ unitary matrix. The sterile neutrinos of the SWSM are much more massive than the active ones. We may safely assume that active-sterile neutrino mixing is negligible (that is, the off-diagonal $3 \times 3$ blocks vanish), and hence the active neutrino mixing matrix is unitary (the $3\times 3$ upper left block of \textbf{U}, ie.~Pontecorvo-Maki-Nakagawa-Sakata matrix). Using these conditions, we can perform the matrix element sums and obtain the simplified expression:
    \bal 
    -\ri eC^L_{Z'\nu\nu} &\approx -\frac{\ri }{2} \left(\frac{e}{\sin \theta_W\cos \theta_W} \sin \theta_Z + (\eta -1) g_z \cos \theta_Z\right)
    .
    \eal
    The other $Z'$-fermion couplings (multiplied by $\ri$ for easier reading) are
    \bal 
    eC^L_{Z'dd} &\approx -\frac{1}{6}  \tan \theta_W   \Big(e \left(3 \cot ^2\theta_W +1\right) \sin \theta_Z +(\eta -1) g_z \cot \theta_W  \cos \theta_Z \Big)\\
    eC^R_{Z'dd} &\approx +\frac{1}{6}   \Big(2 e \tan \theta_W  \sin \theta_Z +(2 \eta -5) g_z \cos \theta_Z \Big)\\
    eC^L_{Z'uu} &\approx -\frac{1}{6}  \tan \theta_W   \Big(e (1-3 \cot ^2\theta_W)   \sin \theta_Z +(\eta -1) g_z \cot \theta_W  \cos \theta_Z \Big)\\
    eC^R_{Z'uu} &\approx -\frac{1}{6}   \Big(4 e \tan \theta_W  \sin \theta_Z +(4 \eta -7) g_z \cos \theta_Z \Big)\\
    eC^L_{Z'ee} &\approx -\frac{1}{2}  \tan \theta_W   \Big(e \left(\cot ^2\theta_W -1\right) \sin \theta_Z -(\eta -1) g_z \cot \theta_W  \cos \theta_Z \Big)\\
    eC^R_{Z'ee} &\approx +\frac{1}{2}   \Big(2 e \tan \theta_W  \sin \theta_Z +(2 \eta -3) g_z \cos \theta_Z \Big)
    \eal 
    Now we may write the Feynman amplitude for virtual $Z'$-mediated $\nu_\ell f \rightarrow \nu_\ell f$-scattering. Then we obtain the NSI couplings derived from the SWSM as
    \be 
    \label{eq:SWSM-NSI}
    \eps^{f,X}(g_z,\eta,\tan \beta) 
    = -\frac{v^2}{2(q^2-M_{Z'}^2)}(eC_{Z'\nu\nu}^L)(eC_{Z'ff}^X)
    \,,
    \ee
    which interpolates smoothly between the limits of heavy or light NSI couplings given by
    \be
    \eps^{f,X}\approx 
    \frac12(eC_{Z'\nu\nu}^L)(eC_{Z'ff}^X)\times 
    \begin{cases}
        \dfrac{v^2}{M_{Z'}^2},\text{ when }M_{Z'}^2 \gg q^2,\\
        \vspace{-0.3cm}\\ 
        -\dfrac{v^2}{q^2},\text{ when } M_{Z'}^2 \ll q^2.
    \end{cases}
    \ee 
    These NSI couplings are flavour universal, hence we have suppressed the corresponding lower indices. Also, flavour is conserved.
    
    The mass of the $Z'$ in Eq.~\eqref{eq:SWSM-NSI} is fixed according to Eq.~(A.14) of Ref~\cite{Iwamoto:2021fup}, reproduced in an equivalent form here:
    \be \label{eq:MZp2}
    M_{Z'}^2(g_z,\eta,\tan\beta) = \frac{g_z^2v^2\tan^2 \beta}{1 + \frac{1}{e}(2 - \eta)g_z\sin\theta_W\cos \theta_W}
    \,,
    \ee 
    with $\tan \beta = w/v$ being the the ratio of the two VEVs. In addition, the mixing angle $\theta_Z$ also depends on the same parameters (see Eq. (A.13) of \cite{Iwamoto:2021fup}),
    \bal 
    \tan 2\theta_Z &= \frac{\brac{1-\frac{\eta}{2}}\frac{g_z\cos \theta_W}{g_L}}{\frac{1}{4}-\brac{\brac{1-\frac{\eta}{2}}^2+\tan^2\beta}\bfrac{g_z\cos\theta_W}{g_L}^2}
    \,.
    \eal
    
    \subsection{Numerical estimates}
    
    Solving the Eq.~\eqref{eq:MZp2} for $g_z$, we obtain for positive $g_z$ that
    \be \label{eq:gz-numeric}
    \bsp
    g_z &= \frac{1}{4 e v^2\tan^2\beta}
    \\ &\times 
    \left(\sqrt{M_{Z'}^2 \left(16 e^2 v^2 \tan ^2\beta +(\eta -2)^2 M_{Z'}^2 \sin ^2\left(2 \theta_W\right)\right)}-(\eta -2) M_{Z'}^2 \sin \left(2 \theta_W\right)\right)\\
    &
    \simeq \frac{3.94\cdot 10^{-6}}{\tan \beta} \times \frac{M_{Z'}}{\text{MeV}}
    \esp
    \ee
    where we substituted $\eta = 0$ and took into account only the leading order contribution. We justify this by noting that in our investigation the dependence of $\eta$ on other parameters is weak and its inclusion is manifested by multiplying the right hand side of Eq.~\eqref{eq:gz-numeric} with a multiplicative factor of $\mathcal{O}(1)$. Similarly,
    \bal
    \theta_Z &\approx (2-\eta)\cos \theta_W\frac{g_z}{g_L} 
    \simeq 1.354(2-\eta)g_z 
    = g_z \times \mathcal{O}(1)
    \,.
    \label{eq:thetaZ-gz}
    \eal 
    Assuming $\theta_Z \ll 1$ (i.e.~super-weak coupling), we can derive the following expressions for NSI couplings:
    \bal 
    \eps^u &\simeq \frac12\bfrac{v}{M_{Z'}}^2\brac{\frac{g_z^2}{12}\brac{-5\eta^2 + 13\eta - 8}+ 0.2355 g_z\theta_Z(1.766-\eta) + 0.0469\theta_Z^2}
    \,,\\
    \eps^d &\simeq \frac12 \bfrac{v}{M_{Z'}}^2\brac{\frac{g_z^2}{12}\brac{\eta^2 - 5\eta + 4}- 0.0626 g_z\theta_Z(1.881+\eta) - 0.0885\theta_Z^2}
    \,,\\
    \eps^e &\simeq \frac12 \bfrac{v}{M_{Z'}}^2\brac{\frac{g_z^2}{4}\brac{3\eta^2 - 7\eta + 4}+ 0.5335 g_z\theta_Z(1.338- \eta) - 0.00536\theta_Z^2}
    \,.
    \eal 
    Scanning over the possible $\eta$, we find
    \be 
    \theta_Z \in [1.820, 2.708] g_z
    \text{~~and~~}
    |\eps^f| \in \mathrm{in}_f \bfrac{v g_z}{M_{Z'}}^2
    \,,
    \ee
    with flavour dependent intervals
    \be
    \mathrm{in}_u = [0.248, 0.402] 
    \,,\qquad
    \mathrm{in}_d = [0.339, 0.651] 
    \,,\qquad
    \mathrm{in}_e = [0.4275, 1.486]
    \,. 
    \ee
    Note that the NSI parameters are not independent of each other, which can be seen by taking the ratio
    of up- and down-type quark NSI in SWSM,
    \be \label{eq:epsratio}
    R=\frac{\eps^u}{\eps^d} = \frac{eC^L_{Z'uu} + eC^R_{Z'uu}}{eC^L_{Z'dd} + eC^R_{Z'dd}} = \frac{e \left(5-3 \cot ^2\theta_W\right) \sin \theta _Z+(5 \eta -8) g_z \cot \theta _W \cos \theta _Z}{e \left(3 \cot ^2\theta _W-1\right) \sin \theta _Z-(\eta -4) g_z \cot \theta _W \cos \theta _Z}
    \,,
    \ee
    from which we can express $\eta$ as
    \be
    \eta = \frac{\frac{e}{g_z} \tan \theta_W \tan\theta_Z \left(3 (R+1) \cot^2 \theta_W-R-5\right)+4 R+8}{R+5} 
    \,.
    \ee
    It turns out that the resulting valid benchmark points are confined to a very narrow region (see the next section).
    
    Finally we remark that assuming a universal bound $\eps_\text{max}$ for the NSI couplings, we may present a simple analytic bound in the $(M_{Z'},g_z)$ plane, namely
    \be 
    g_z < \sqrt{\eps_\text{max}}\bfrac{M_{Z'}}{v} \times \mathcal{O}(1)
    \,.
    \ee 
    
    \section{Results}
    
    Our results are two-fold. First we present constraints on the parameters of the SWSM and also on the NSI parameters originating from the SWSM. Next we discuss our predictions for those NSI couplings.
    
    \subsection{Free parameters and constraints}
    
    The NSI couplings depend on the gauge sector parameters 
    $g_z$, $\eta$ and on $\tan \beta$, which we choose as free parameters in the model.
    For the neutrino masses we consider, we may assume that the PMNS matrix is unitary, since nonunitary effects contributing to the NSI are negligible \cite{Karkkainen:2021tbh}.
    
\begin{figure}[t]
        \centering
        \includegraphics[width=\linewidth]{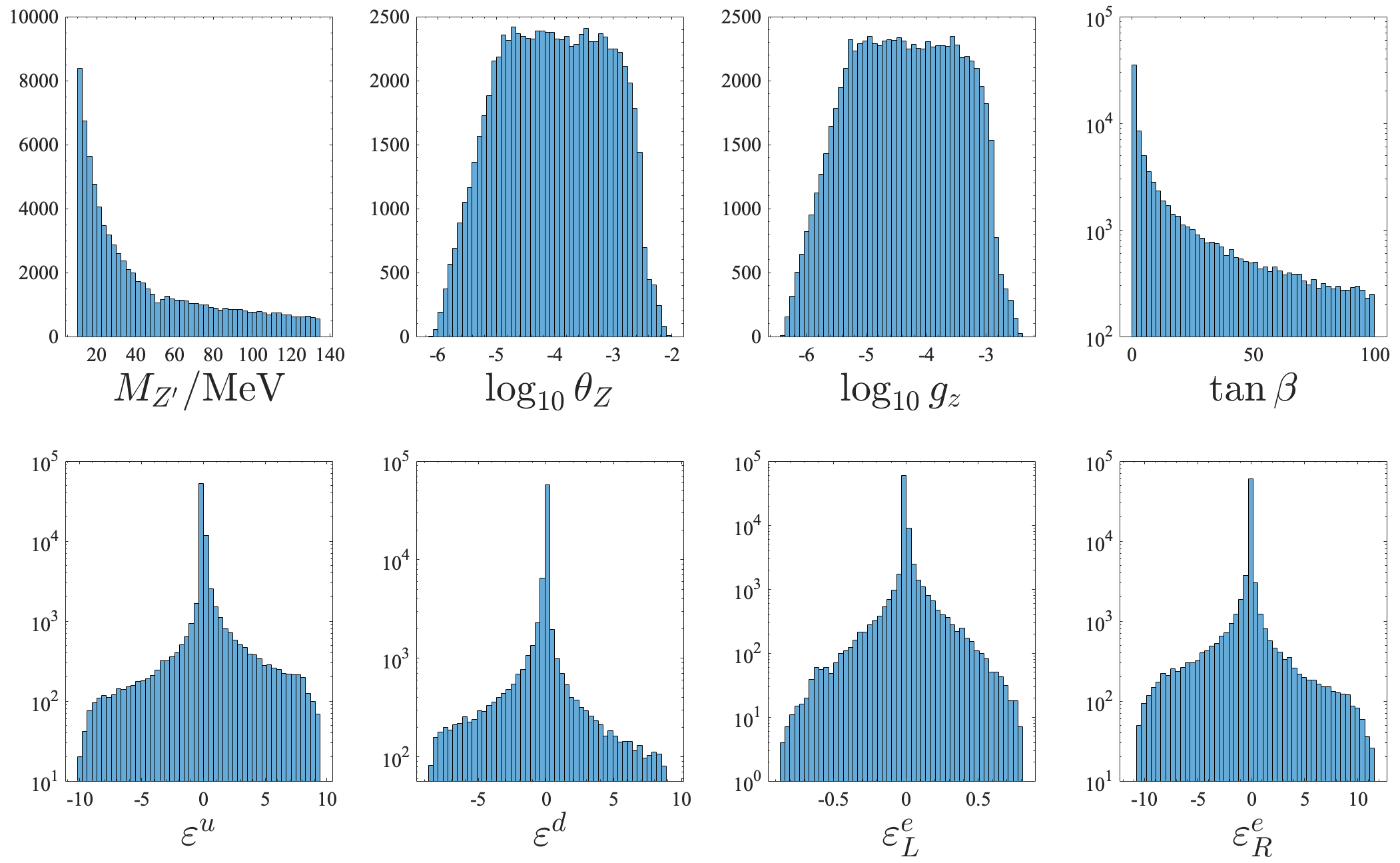}
        \caption{
            \label{fig:histograms}
            Histograms (containing 50 bins) of the scan (with total number of points $N=10^6$) corresponding to $M_{Z'}$, $\log_{10}\theta_Z$, $\log_{10}g_z$, $\tan\beta$, $\eps^u$, $\eps^d$, $\eps_L^e$ and $\eps^e_R$. Note that the first three of the histograms have linear, while the last five ones have logarithmic vertical axis.}
    \end{figure}

    We scanned the $(\log_{10}\tan \beta,\log_{10}|g_z|,\eta)$ right rectangular prism by a uniformly distributed random sampling in $[-2,2]\times [-10,0] \times [0,0.656]$ to determine the 
    region consistent with current bound on isotropic NSI couplings, derived in Sec.~2.3. Larger values of $\tan\beta$ are possible in principle, but in such cases the new scalar sector decouples almost completely, hence remains inaccessible. Also values $\tan\beta \gtrsim 100$ are disfavored by the overproduction of dark matter if the SWSM is to explain the origin of dark matter energy density observed in the Universe \cite{Seller:2022noz}. We used the $2\sigma$ limits for the NSI couplings as given in Table~\ref{tab:NSIbounds4}. We present the allowed values in histograms in Fig.~\ref{fig:histograms} and in Table~\ref{tab:stats}. We see that the model prefers small values of $M_{Z'}$ and $\tan \beta$. The distribution of $g_z$ (hence also $\theta_Z$) is fairly flat within the allowed range $g_z \in 5 \cdot [10^{-6}, 10^{-4}]$ (approximately), with the full allowed range being somewhat larger.
    We note that the average value (or also the median) of the asymmetric $\eps^u$ and $\eps^d$ distributions are positive and negative, since they are skewed to the respective values.
    \begin{table}[!ht]
        \centering
        \begin{tabular}{|c|c|c|c|}\hline 
            \rule{0pt}{3ex}\textbf{Parameter}  & \textbf{Scan range } & \textbf{BP range ($2\sigma$)} & \textbf{BP range ($1\sigma)$}\\\hline 
            $\eta$              &[0,0.656] & [0,0.656]  & [0,0.656] \bigstrut \\\hline 
            $\tan \beta$        &[0.01,100]& [0.02,100]& [0.03,100] \bigstrut \\\hline 
            $\log_{10}g_z$      &[$-10$,1] & [$-6.38,-2.31$] & [$-6.38,-2.41$] \bigstrut \\\hline 
            $M_{Z'}/$MeV        &[10,135]  & [10,135]   & [10,135] \bigstrut \\\hline 
            $\log_{10}\theta_Z$ &--        & [$-6.09,-1.94$] & [$-6.09,-2.05$] \bigstrut \\\hline 
            $\eps^u$            &[$-17.25,17.16$]&[$-17.25,17.16$]& $[-10.00,9.42]$ \bigstrut \\\hline 
            $\eps^d$            &[$-15.31,16.05$]&[$-15.31,16.05$]& $[-8.606,9.221]$ \bigstrut \\\hline 
            $\eps^e_L$          &--&$[-1.504,1.462]$&$[-0.856,0.808]$ \bigstrut \\\hline 
            $\eps^e_R$          &--&$[-19.87,19.84]$&$[-10.91,11.55]$ \bigstrut \\\hline 
        \end{tabular}
        \caption{Scan and benchmark point ranges corresponding 2 and $1\sigma$ allowed regions of COHERENT experiment.
            \label{tab:stats}}
    \end{table}
    
    \subsection{Predictions}
    \label{sec:SWSM}
    
\begin{figure}[t]
        \centering
        \includegraphics[width=0.49\linewidth]{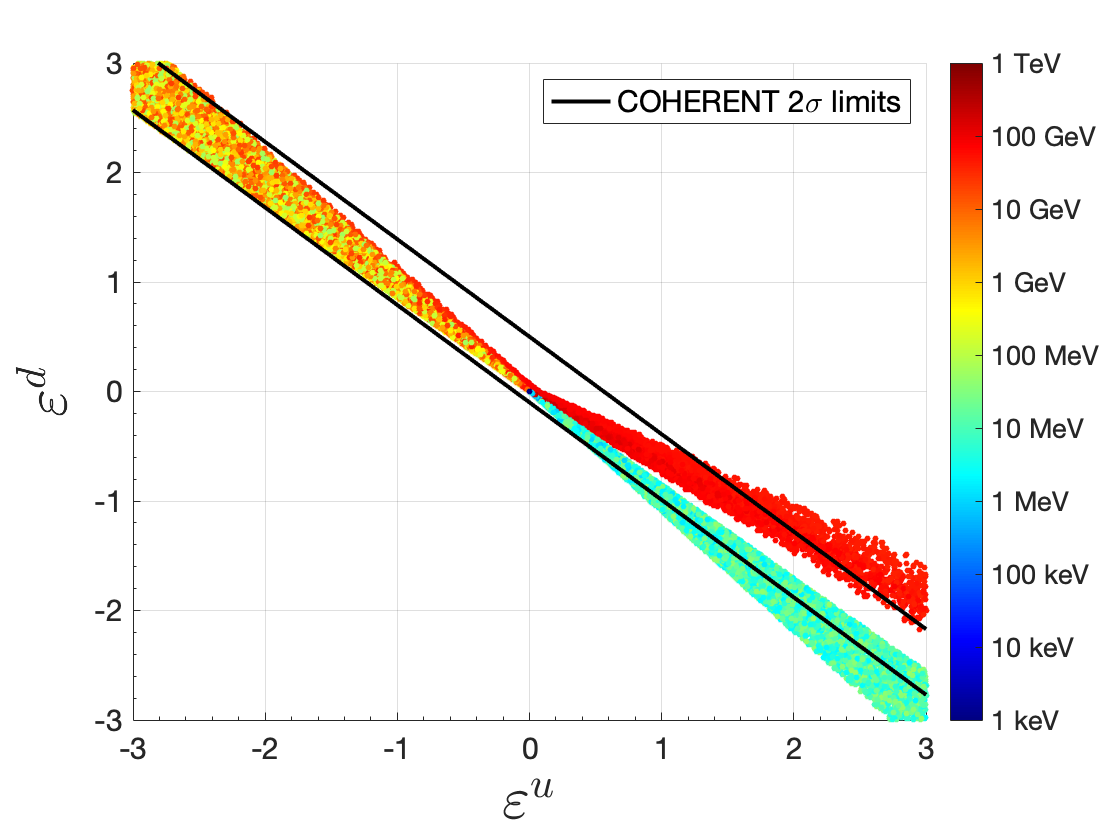}
        \includegraphics[width=0.49\linewidth]{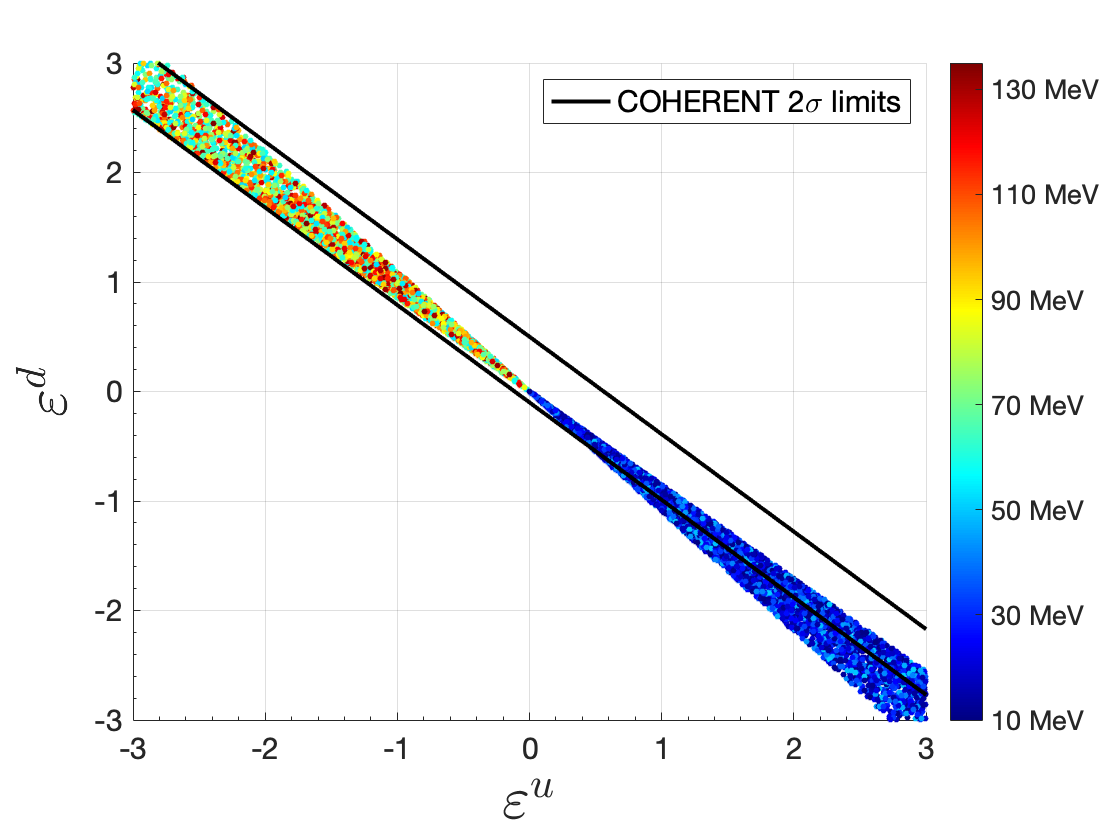}
        \caption{Left: Available parameter space in $(\eps^u,\eps^d)$ plane corresponding to the scan ranges in Table~\ref{tab:stats} except that for the mass of the $Z'$, for which $M_{Z'} \in [1,10^9]$\,keV. Right: benchmark points consistent with SWSM freeze-out dark matter scenario. 
            \label{fig:mzp-heat}}
    \end{figure}

    The NSI couplings $\eps^u$ and $\eps^d$ derived from the SWSM are anticorrelated, as can be seen on Fig.~\ref{fig:mzp-heat} obtained using those in Eq.~\eqref{eq:SWSM-NSI} with $q^2 \simeq (51\,\mathrm{MeV})^2$ as the characteristic energy transfer squared in the COHERENT experiment.
    The region between black lines is consistent with the $2\sigma$ bounds from COHERENT. The data points are coloured according to the mass of the $Z'$. Three distinct $Z'$ mass regions emerge. 
    In the lower right sector two clearly different $Z'$ mass regions can be identified: light (turquoise) and heavy (red) areas. 
    The region with red colour in the left plot is inconsistent SWSM freeze-out dark matter scenario, which requires that the mass of the $Z'$ boson falls into the (10--135)\,MeV mass range \cite{Iwamoto:2021fup}. Restricting our scan to this constrained region, shown on the right plot, reveals additional predictions: if $q \lesssim M_{Z'} \leq m_\pi$, then $\eps_u < 0 < \eps_d$ but if 10\,MeV $\leq M_{Z'} \lesssim q$, then $\eps_d < 0 < \eps_u$.
    
    \begin{figure}[ht!]
        \centering
        \includegraphics[width=0.48\linewidth]{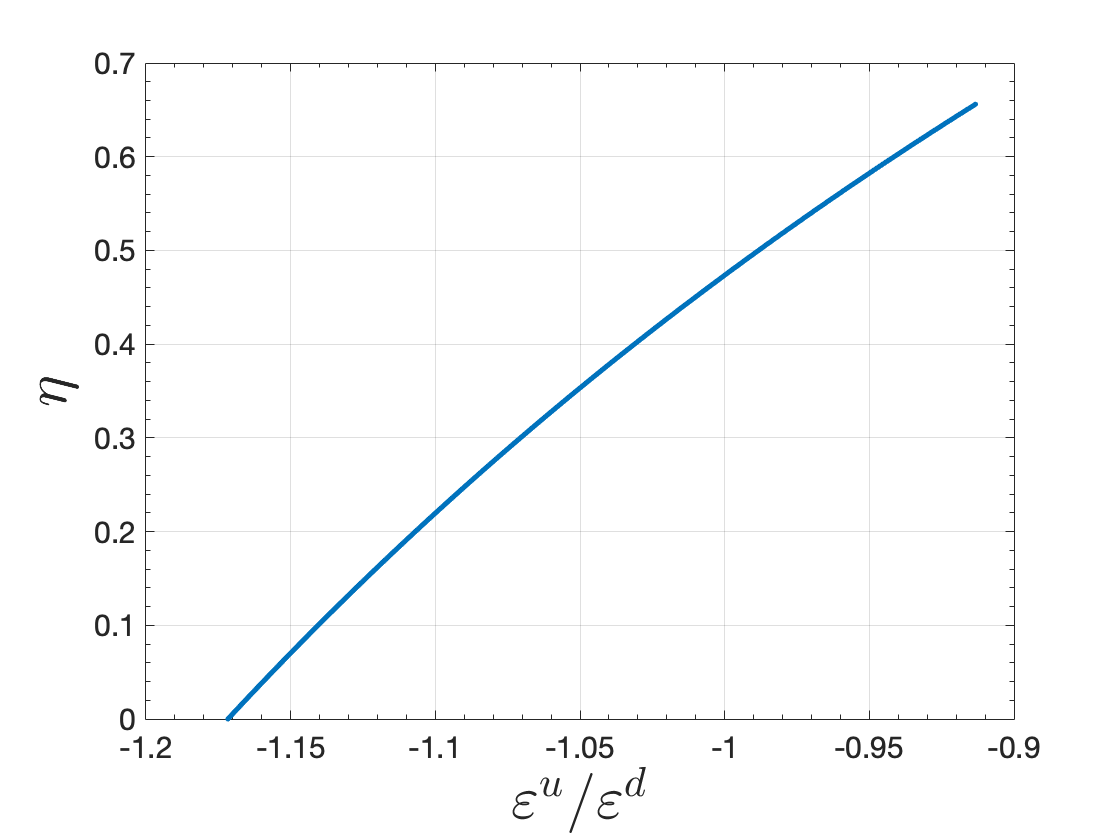}
        \hfill
        \includegraphics[width=0.48\linewidth]{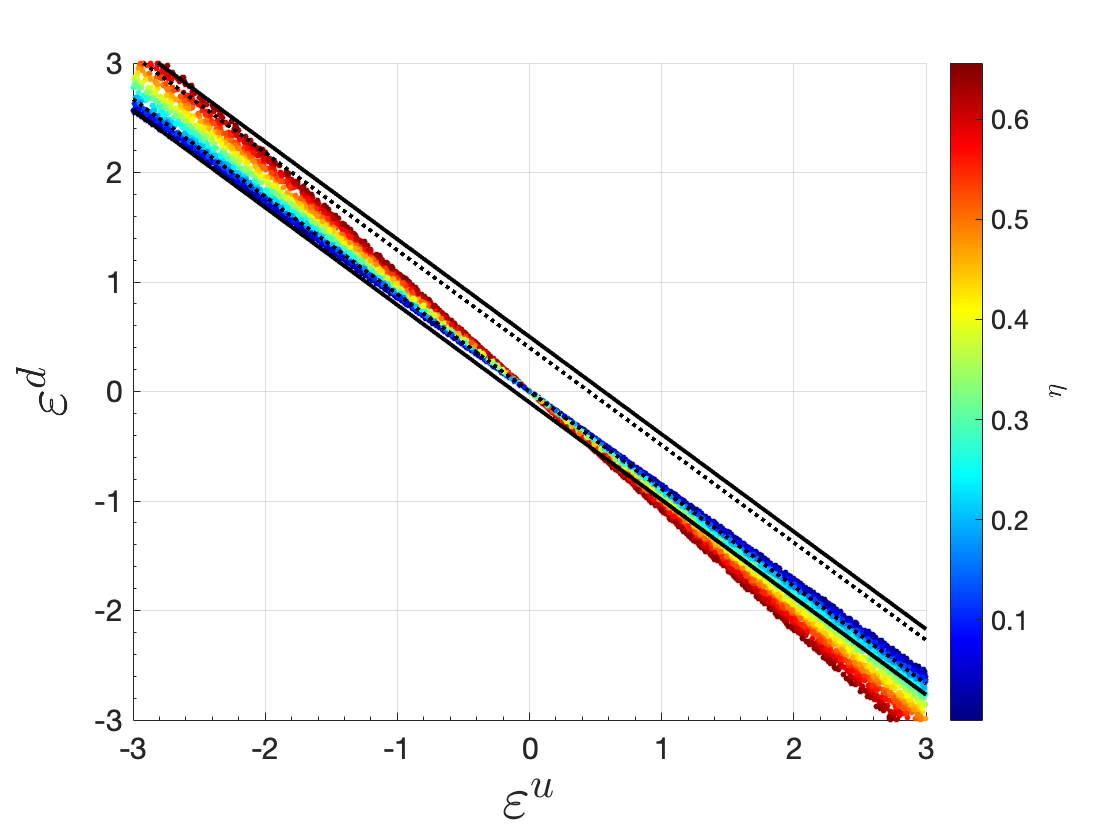}
        \caption{Left: The $\eta$ parameter as a function of $\eps^u/\eps^d$. Right: As in Fig.~\ref{fig:mzp-heat} right panel, but the data points are coloured according to $\eta$, which corresponds to azimuthal angle in $(\eps^u,\eps^d)$ plane. 
            \label{fig:epsratio}}
    \end{figure}
    
    In the left panel of Fig.~\ref{fig:epsratio} we can see that the parameter $\eta$ is almost a linear function of the ratio $(\eps^u/\eps^d)$ as one expects based on the discussion after Eq.~\eqref{eq:epsratio}. This information is visualized as a heat map in the right panel of Fig.~\ref{fig:epsratio}, which shows that the COHERENT limits are compatible with $\epsilon^u > 0$ at the $2\sigma$ confidence level only for $\eta \lesssim 0.3$ at the electroweak scale. 
    The region between dotted lines corresponds to $1\sigma$ bounds from COHERENT.
    
\begin{figure}[t]
        \centering
        \includegraphics[width=\linewidth]{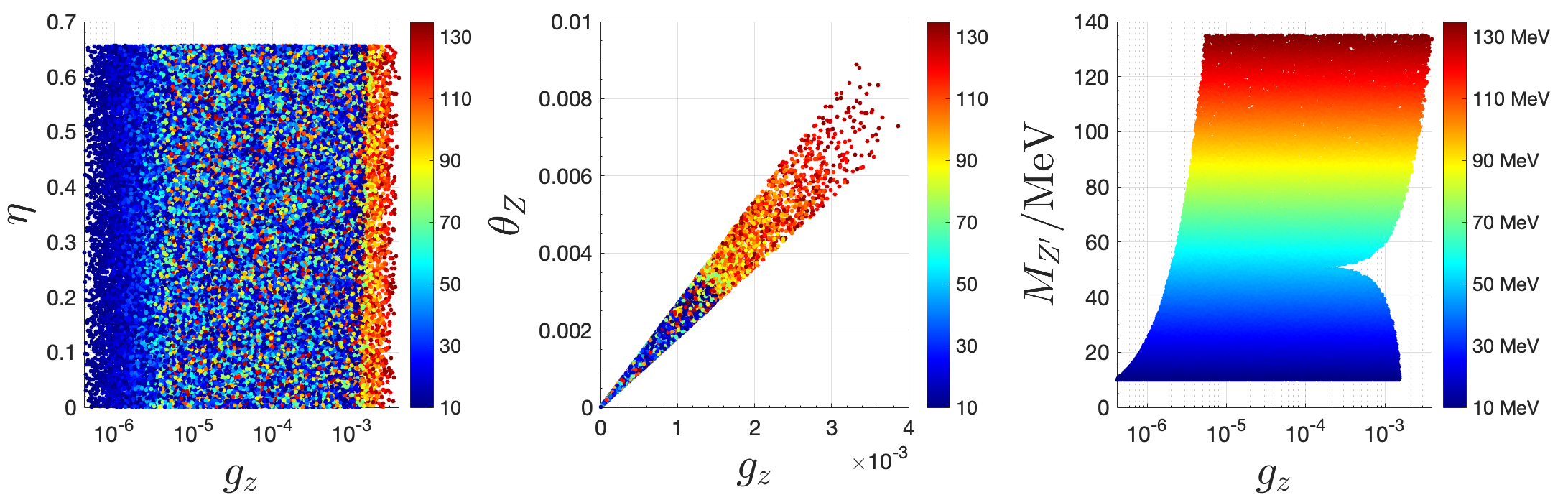}
        \caption{
            \label{fig:mzp-heat-sweak} Benchmark points in $(g_z,X)$ planes, with $X=\eta$, $\theta_Z$ and $M_{Z'}$. The colour corresponds to $M_{Z'}$ in MeV units.}
    \end{figure}

\begin{figure}[ht!]
        \centering
        \includegraphics[width=0.7\linewidth]{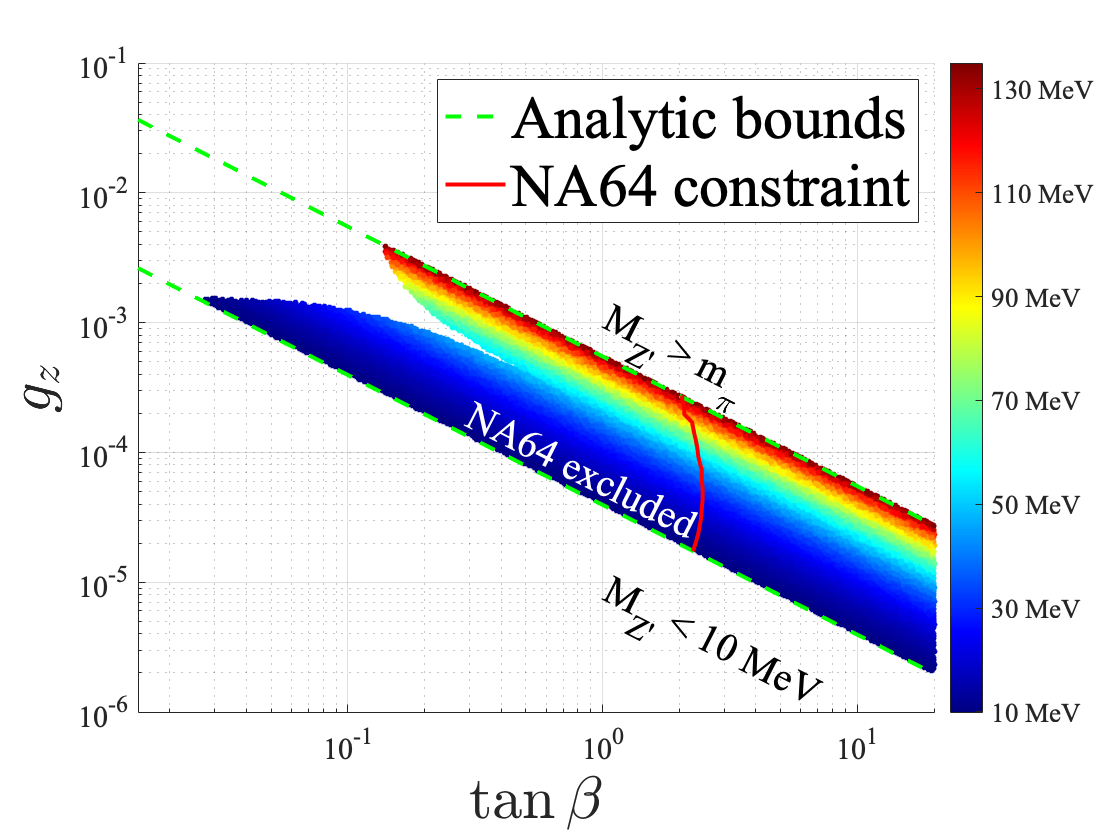}
        \caption{
            \label{fig:gz-tb-2} Available parameter space in $(\tan \beta, g_z)$ plane, where colour corresponds to $M_{Z'}$. Lower and upper analytic bounds for $M_{Z'}$ obtained from Eq.~\eqref{eq:gz-numeric} using the allowed region for $M_{Z'}$ assuming the freeze-out dark matter scenario of SWSM \cite{Iwamoto:2021fup}. The solid red line represents the NA64 constraint on direct dark photon search.}
    \end{figure}

    We present additional benchmark points (BPs) in Fig.~\ref{fig:mzp-heat-sweak} over the $(g_z,X)$ planes ($X=\eta$, $\theta_Z$ and $M_{Z'}$) as heat maps depending on the mass of the $Z'$. All these plots are relevant in the context of explaining dark matter within the SWSM. The BPs do not exhibit any particular dependence on the parameter $\eta$ representing the kinetic mixing. The second plot visualizes precisely the approximate relation in Eq.~\eqref{eq:thetaZ-gz}.
    We show the available parameter space in $(\tan\beta, g_z)$ plane separately in Fig.~\ref{fig:gz-tb-2} where we present approximate analytic bounds superimposed (green dashes). In addition we added the NA64 constraint obtained by searching for dark photons, identified here with the $Z'$ (red solid curve) \cite{Seller:2022noz}.
    For $\tan \beta$ we find the lower bounds corresponding to NA64 slightly depending on the value of the coupling $g_z$. The gauge coupling is constrained to $2\sigma$ confidence interval between $4.17 \cdot 10^{-6}$ and $4.90 \cdot 10^{-3}$, where the lower and upper bounds correspond to $M_{Z'} > 10$ MeV and $M_{Z'} < m_\pi$.
    We see that the mass of the $Z'$ does not significantly affect this $\tan \beta$ bound, but the favoured values of $\theta_Z$ increase with $M_{Z'}$ (see Fig.~\ref{fig:mzp-heat-sweak}).
    
\begin{figure}[t]
        \centering
        \includegraphics[width=0.8\linewidth]{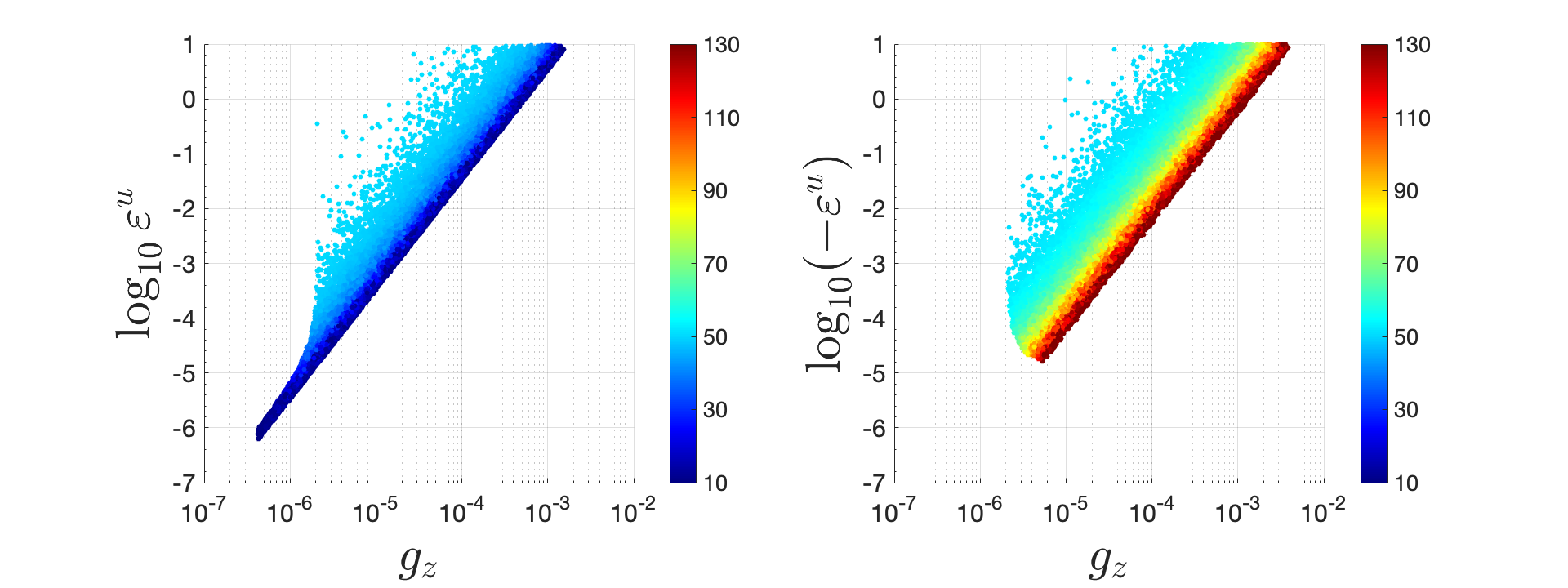}
        \includegraphics[width=0.8\linewidth]{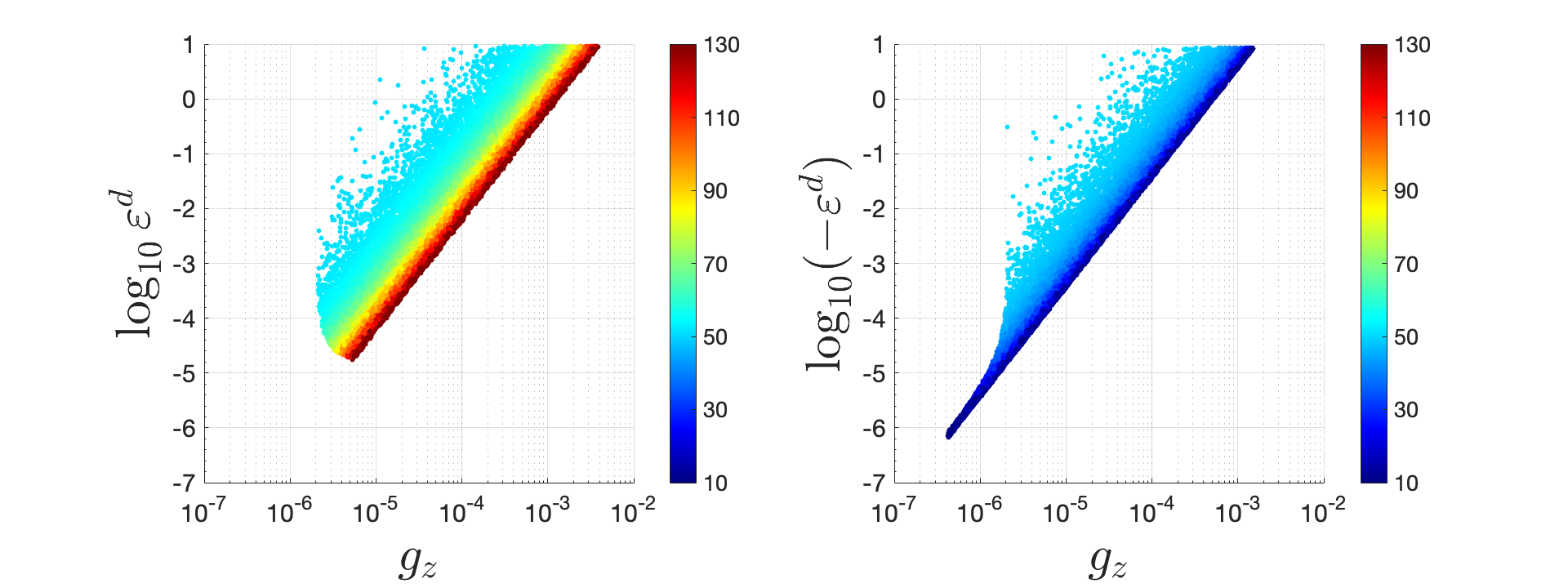}
        \caption{
            \label{fig:mzp-heat-eps} Benchmark points in $(g_z,\log_{10}(\pm\eps^u))$ and $(g_z,\log_{10}(\pm\eps^d))$ planes. Note that different signs of the NSI parameters correspond to two regions of parameter space:  $M_{Z'} < 51$ MeV and $M_{Z'} > 51$ MeV. We separated the cases corresponding to positive and negative values of $\eps^u$ and $\eps^d$. The colour corresponds to $M_{Z'}$.}
    \end{figure}

    While large NSI couplings are still allowed, according to Fig.~\ref{fig:histograms} for the benchmark point distributions small couplings are favoured in $\eps^u,\:\eps^d,\:\eps^e_L$ and $\eps^e_R$. The corresponding BPs are shown in Fig.~\ref{fig:mzp-heat-eps}.
    
    \section{Conclusions and future prospects}

    We have considered an exciting possibility for NSI, which escapes the high-energy experimental constraints and detection by neutrino oscillation experiments. Former experiments are unable to probe the interactions with a light mediator, while flavour-universal couplings between the mediator and a neutrino are manifested as an irrelevant phase factor in neutrino oscillation Hamiltonian. In the presence of sterile neutrinos the factor does not disappear, but is suppressed \cite{Karkkainen:2021cmo}.
    
    The only viable avenue to probe flavour-universal light NSI couplings is then to consider CE$\nu$NS. We derived the bounds for flavour-universal NSI both in light and heavy mediator case, and found that large NSI couplings $(\eps \simeq 10)$ are allowed for the light NSI scenario, while $\eps \lesssim 10^{-2}$ for the heavy case. 
    
    We then considered a specific model, the super-weak extension of the standard model. We obtained the NSI couplings in the SWSM, which allowed us to investigate the parameter space of the SWSM as allowed by the existing constraints of CE$\nu$NS on the NSI parameters.
    We found that in this range the model prefers small values for the mass of the new gauge boson and also for the ratio $w/v$ of the VEVs. The kinetic mixing parameter is weakly constrained, but we found that its possible values are compatible with $\varepsilon^u/\varepsilon^d \in [-1.17,-0.92]$. This ratio of NSI strengths is a testable prediction of the SWSM.
    If we added the constraint set by the NA64 experiment on the mass of dark photon, we could constrain further
    the viable parameter space to $\tan \beta \gtrsim 2$ and $g_z \sim 10^{-6}-10^{-3}$. 
    
    Our study demonstrated that even low-energy experiments have significant potential on constraining new physics discovery. Both higher-intensity and higher-energy experiments are needed for the progressive discovery of light and heavy NSI interactions. While the limits from CE$\nu$NS are quite loose at present, their expected improvement at the ESS \cite{Baxter:2019mcx} will constrain the parameter space of the SWSM severely.
    
    
    %
    
\end{document}